\documentclass[aps,pra,twocolumn,10pt,superscriptaddress,longbibliograhy,floatfix]{revtex4-2}

\usepackage[T1]{fontenc}
\usepackage{natbib}
\usepackage{indentfirst}
\usepackage{listings}
\usepackage{babel}
\usepackage{braket}
\usepackage{amsmath}
\usepackage{amssymb}
\usepackage{amsthm}
\usepackage{graphicx}
\usepackage{float}
\usepackage{array, makecell}
\usepackage[dvipsnames]{xcolor}
\usepackage{bbold}
\usepackage{placeins}
\usepackage{xcolor}
\DeclareMathOperator{\Tr}{Tr}
\usepackage[title]{appendix}
\usepackage[export]{adjustbox}

\usepackage[colorlinks=true,linkcolor=blue,citecolor=blue,urlcolor=blue]{hyperref}

\begin{document}
\title{Arithmetic sequences as quantum states}

\author{Ruge Lin}
\affiliation{Quantum Research Centre, Technology Innovation Institute,  United Arab Emirates.}
\affiliation{Departament de F\'isica Qu\`antica i Astrof\'isica and Institut de Ci\`encies del Cosmos, Universitat de Barcelona, Spain.}

\author{Germ\'an Sierra}
\affiliation{Instituto de F\'isica Te\'orica UAM-CSIC, Universidad Aut\'onoma de Madrid, Spain.}

\author{Jos\'e I. Latorre}
\affiliation{Quantum Research Centre, Technology Innovation Institute,  United Arab Emirates.}
\affiliation{Centre for Quantum Technologies, National University of Singapore, Singapore.}

\begin{abstract}

We consider arithmetic sequences, here defined as ordered lists of positive integers. Any such a sequence can be cast onto a quantum state, enabling the quantification of its `surprise' through von Neumann entropy. We identify typical sequences that maximize entanglement entropy across all bipartitions and derive an analytical approximation as a function of the sequence length. This quantum-inspired approach offers a novel perspective for analyzing randomness in arithmetic sequences.

\end{abstract}

\maketitle

\def\barray{\begin{eqnarray}}
\def\earray{\end{eqnarray}}
\def\beq{\begin{equation}}
\def\eeq{\end{equation}}

\section{Introduction}

Arithmetic questions can be numerically explored with classical computers. This approach is limited by the size of memory and processing power of our present technology, as many problems require the ability to work in the orders of magnitudes away from any foreseeable classical technology. 


An example of an arithmetic question that is difficult
to analyze numerically is to compute the difference between the prime counting function $\pi(x)$ and the logarithmic integral function $li(x)$, which estimates 
$\pi$(x) asymptotically. Littlewood showed in 1914 that their difference changes sign infinitely often as $x$ increases, without indicating the locations of these changes \cite{littlewood1914sur}. Skewes later estimated an extremely large upper bound for the first sign change \cite{skewes1933on}, which has since been refined to occur before $1.397 \times 10^{316}$,
far beyond the reach of classical computation \cite{saouter2010sharp,stoll2011impact}. A natural question arises: could quantum computation offer new insights into understanding these sign changes?

Translating arithmetic problems into a quantum framework is the initial step toward exploring arithmetic questions through quantum computation. Quantum algorithms could then be designed to uncover the properties of these arithmetic problems. Prior studies have revealed intriguing links between prime numbers, entanglement, and their foundational roles in both arithmetic and quantum mechanics \cite{latorre2013quantum, latorre2014there, garcia2020prime}. Prime sequences display quantum-like properties, showing high entanglement levels that relate to their unpredictability. Furthermore, the Riemann hypothesis, which addresses the distribution of primes, might be observed as a physical phenomenon in many-body physics \cite{mussardo2018generalized, sierra2019riemann}. This suggests that if number theory has physical implications, entanglement could potentially exhibit intrinsic arithmetic characteristics.

In this article, we delve into the quantum properties of general arithmetic sequences, with a specific focus on the von Neumann entropy of natural bipartitions. We begin by revisiting the mapping of arithmetic sequences to quantum states. Subsequently, we investigate the average entanglement of arithmetic sequences as a benchmark and provide analytical interpretations for our observations. Additionally, we extend our exploration to include the concept of the unions of $k$-almost prime numbers and show the expected preservation of entanglement when switching from position space to momentum space.

\section{Arithmetic Sequences as Quantum States}

In Number Theory, a sequence is an ordered list of integers. Here we consider finite sequences consisting of $M$ distinct positive integers within the interval $[0, N-1]$, 
\begin{equation}
\mathbb{A}_{N,M} =\{a_1,\ldots,a_{M} \}, \quad 0 \leq a_1 <  \dots < a_{M} \leq N -1 \, ,
\label{eq_seq}
\end{equation}
with $M \in [1, N]$. The number of such sequences is given by the binomial coefficient 
$\binom{N}{M}$. 

Let us now show how any arithmetic sequence can be cast on a quantum state. The computational basis for the quantum register consists of the states 
$\ket{\mathbf{x}}=\ket{x_{n-1}}\otimes\dots\otimes\ket{x_1}\otimes\ket{x_0}
$, 
where $\{ x_j  \}_{j=0}^{n-1}$ are the binary digits of the integer ${\bf x}$, that is 
${\bf x} = x_0 2^0+ x_1 2^1+ \dots + x_{n-1} 2^{n-1} \in [0, 2^n-1]$. 
Using the above basis, we can associate a state $|a_j\rangle$ to each element $a_j$ of the sequence in Eq. \eqref{eq_seq} 
provided that $N = 2^n$. Similarly, we can associate a quantum state with the sequence 
$\mathbb{A}_{N, M}$ by creating a uniform and normalized superposition of all states $|a_j\rangle$,  
\begin{equation}
\ket{\mathbb{A}_{N,M}} =\frac{1}{\sqrt{M}}\sum_{j=1}^{M} \ket{a_j}, \quad a_j \in  \mathbb{A}_{N,M} \, .
\label{eq_seqket}
\end{equation}

An example of a trivial sequence is   $\mathbb{A}_{N, N}=\{0,1,2,\ldots, N-1\}$, which can be cast as the following quantum state
\begin{equation}
\ket{\mathbb{A}_{N,N}}= 
\frac{1}{\sqrt{N}}\sum_{a=0}^{N-1} \ket{a}=\frac{1}{\sqrt{N}}(\ket{0}+\ket{1})^{\otimes n} \, ,
\label{eq_allbasis}
\end{equation}
consisting of the linear superposition of all states of the computational basis. This state is commonly used in quantum algorithms to initialize a register with all possible elements of the computational basis. 

\section{Entanglement measure}

Once an arithmetic sequence is encoded onto a quantum register, it is natural to examine the amount of entanglement within the state. Several metrics can describe quantum correlations. 
Here, we will consider the von Neumann entropy of the reduced density matrix for half of the qubits in the register. This measure has been highly effective in characterizing quantum phase transitions in condensed matter physics. It also serves as a natural quantification of surprise and has been explored in sequences involving prime numbers.

Let us review how the entropy measure is computed from a ket $\ket{\mathbb{A}_{N, M}}$  which belongs to ${\cal H}_n= (\mathbb{C}^2)^{\otimes n}$. If $n$ is an even number, we can split ${\cal H}_n$  in the tensor product ${\cal H}_{n/2}^L \otimes {\cal H}_{n/2}^R$, corresponding to the partition of $n$ qubits into their left and right-handed parts with $n/2$ qubits each. Tracing over the left Hilbert space, one gets the reduced density matrix of a state $\ket{\mathbb{A}_{N, M}} \in {\cal H}_n$
\begin{equation}
\rho_{\mathbb{A}_{N, M}}=\Tr_{{\cal H}^L_{n/2}} \ket{\mathbb{A}_{N, M}} \bra{\mathbb{A}_{N, M}} \, .
\label{eq_rho}
\end{equation}

The von Neumann entropy of this density matrix is defined by
\begin{equation}
S_{\mathbb{A}_{N, M}}  =-  \Tr_{\mathcal{H}^R_{n/2}} \rho_{\mathbb{A}_{N, M}} \log_2 \rho_{\mathbb{A}_{N, M}} \, ,
\end{equation}
and can be computed as 
\begin{equation}
S_{\mathbb{A}_{N, M}}=-\sum_{j=0}^{\sqrt{N} -1} \lambda_j \log_2 \lambda_j \, ,
\label{eq_SANM}
\end{equation}
where $\lambda_j \geq 0$ are the  eigenvalues of $\rho_{\mathbb{A}_{N, M}}$ and satisfy a normalization relation $\sum_j \lambda_j =1$. 
Note that the eigenvalues $\lambda_j$ obtained by tracing out the left half of the qubits are identical to those resulting from tracing out the right half. Therefore, the amount of surprise for any left-right partition is symmetric.

As a simple example, the reduced density matrix of the full superposition state in Eq. \eqref{eq_allbasis} is given by
\begin{equation}
\rho_{\mathbb{A}_{N,N}}=\ket{\mathbb{A}_{\sqrt{N}, \sqrt{N}}}   \bra{\mathbb{A}_{\sqrt{N} \, , \sqrt{N}}} \, , 
\end{equation}
which has a single eigenvalue equal to 1 and, therefore 
\begin{equation}
S_{\mathbb{A}_{N,N}}=0 \, ,
\end{equation}
consistent with the fact that we are considering a product state. 

All sequences with only one element, meaning $M=1$, have zero entanglement  entropy because the corresponding state in Eq. \eqref{eq_seqket} consists of just a single state, to be precise, a product state, in the computational basis, 
\begin{equation}
S_{\mathbb{A}_{N,1}} = 0 , \quad \forall \mathbb{A}_{N,1} \, . 
\end{equation}

For the left-right partition introduced above, a state with maximal entropy is represented by
the tensor product of $n/2$ Bell states across the center of the quantum register, 
\begin{equation}
\ket{\mathbb{A}_{N,\sqrt{N}}}=\ket{\psi^+}_{n-1, 0}\otimes\ket{\psi^+}_{n-2,1}\otimes\dots\otimes\ket{\psi^+}_{\frac{n}{2}, \frac{n}{2}-1} \, , 
\end{equation}
where
\begin{equation}
\ket{\psi^+}_{i,j}=\frac{1}{\sqrt{2}}\left(\ket{0}_i\otimes\ket{1}_j+\ket{1}_i\otimes\ket{0}_j\right) \, , 
\end{equation}
and it has the entanglement entropy
\begin{equation}
S_{\mathbb{A}_{N,\sqrt{N}}}=n/2 \, .
\label{eq_rainbow}
\end{equation}
This state was introduced in Ref. \cite{vitagliano2010, ramirez2014}, 
where it was termed the {\em rainbow} state due to its geometric representation, featuring concentric arcs that connect the left and right parts of the quantum register.

\section{Average entanglement of arithmetic sequences}

We now address the question of the average entanglement present in the quantum states associated with arithmetic sequences. Compared to Haar random states \cite{biswas2021inhibition,mele2024introduction}, states derived from arithmetic sequences lack many possible superpositions, and those that do exist have uniformly weighted coefficients.  The states under consideration have coefficients of either 1 or 0 for each element, indicating the presence of specific integers in the sequence with proper normalization. These states do not involve negative signs or complex numbers, which can potentially limit or enhance the amount of entanglement.

We first examine numerically the entanglement properties of arithmetic sequences. The analysis of random sequences reveals a key property: the entanglement present in a state associated with a sequence primarily depends on the number of terms in that state. Since each term corresponds to an integer in the arithmetic sequence and the states 
are of finite size $N$, the entanglement is observed to depend on $M$, which is the length of the finite sequence. More precisely, entanglement only reaches its possible maximum for sequences with a specific length.

\begin{figure}
\centering
\includegraphics[scale=0.65]{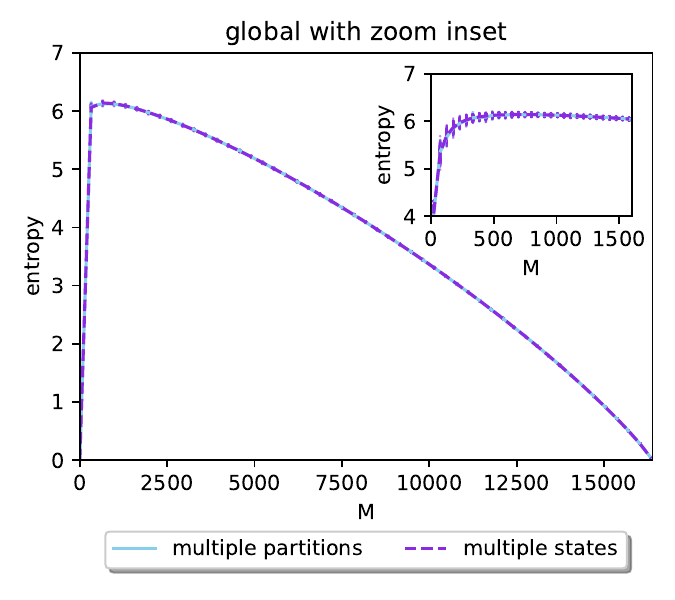}
\caption{The von Neumann entanglement entropy for arithmetic sequences as $n=14$. For each $M$, the data points in blue are generated using one single random arithmetic sequence and averaged over multiple random bi-partitions; the data points in purple are generated using multiple random arithmetic sequences and the natural bi-partitions. The code for every figure in this article is on GitHub \cite{github}.}
\label{fig_Concentration}
\end{figure}

We show in Fig. \ref{fig_Concentration} the dependence of entanglement as a function of the length M for a random sequence of $n$ qubits. It is numerically verified that, for given $N$ and $M$, the entanglement of multiple random bipartitions of a particular random state, as well as the natural bipartitions of many random states, converge to the same value. Moreover, the average entropy as a function of $M$ forms a smooth curve for $M \in [1, N]$, which suggests the possibility of finding an analytical approximation.

The dependence of the von Neumann entropy on $M$ can also be extended to other entanglement measures, such as the R\'{e}nyi entropy, as indicated in Appendix \ref{app_renyi}.

\section{Maximally entangled sequences}

Let us investigate which states with a fixed value of $N$ exhibit the highest von Neumann entropy across all their bi-partitions.

States which are maximally entangled in all their partitions, known as AME (Absolutely Maximally Entangled) states \cite{helwig2012absolute, goyeneche2014genuinely, goyeneche2015absolutely} or perfect states  \cite{pastawski2015holographic}, do not exist for $n \ge 7$ qubits \cite{huber2017}. We will here adopt a statistical approach to search for states that achieve the highest average entropy. 

This search is carried out in two steps: i) For fixed values of $N$ and $M$, select a random subset of sequences $\mathbb{A}_{N,M}$ and compute the average entropy using Eq. (\ref{eq_SANM}) for the corresponding states, defined as:
\begin{equation}
E_{N,M}=\langle{S}_{\mathbb{A}_{N,M}}\rangle \, .
\label{eq_ENM}
\end{equation}
ii) Determine the value of $M$ that maximizes the expression in Eq. \eqref{eq_ENM}:
\begin{equation}
M_n = {\rm arg \; max}_M  \;  E_{N,M}  \, .
\end{equation}
The maximum entropy value is then given by:
\begin{equation}
E_n = E_{N,M_n}  \, .
\end{equation}

To identify the $M_n$ value that results in maximal entanglement, we perform the scaling multiple times around the peak for $M$, using more samples and a quadratic fit to pinpoint the exact value. The numerical simulation results for $M_n$ and $E_n$ for up to 30 qubits are presented in Table I. In particular, almost all sequences with the value $M_n$ exhibit extremely high entanglement across different partitions, as detailed in Appendix \ref{app_partitions}.

\begin{table}
\centering
\vspace{1cm}
\begin{tabular}{lcc}
\hline
$n$ & $M_n$ & $E_n$ \\
\hline
10 & 107 & 4.072 \\
12 & 276 & 5.108 \\
14 & 716 & 6.143 \\
16 & 1873 & 7.176 \\
18 & 4934 & 8.196 \\
20 & 13091 & 9.215 \\
22 & 34771 & 10.231 \\
24 & 93018 & 11.242 \\
26 & 250660 & 12.251 \\
28 & 672556 & 13.258 \\
30 & 1836685 & 14.263 \\
\hline
\end{tabular}
\caption{Values of $M_n$ and $E_n$.}
\label{tab:values}
\end{table}

The linear regression for $E_n$ and $\log_2 M_n$ presented in Table I yields the following results, as shown in FIG. \ref{fig_Linear}
\begin{equation}
\begin{aligned}
E_n & \simeq 0.509 n - 0.990 \, ,\\
\log_2 M_n & \simeq 0.703 n - 0.357 \, .
\end{aligned}
\label{eq_linear}
\end{equation}

The second relationship offers a reliable prediction of $M_n$ for systems with slightly more than 30 qubits. While the coefficient of $n$ for $E_n$ cannot exceed 0.5 \cite{page1993average}, the fit indicates that it approaches this limit as $n$ grows large ($n \gg 1$). Later, we will present an approximation that supports this conjecture.

\section{Asymptotic analysis}

For the arithmetic sequences we are studying, the spectrum of the von Neumann entropy can be divided into two distinct parts: a dominant largest eigenvalue $\lambda_0$ and an almost continuum spectrum that is well separated from $\lambda_0$, as shown in FIG. \ref{fig_Spectral}.

The entries of the density matrix $\rho_{\mathbb{A}_{N, M}}$ are correlated, and can be represented as $\Omega^{\dagger}\Omega/M$, while $\Omega$ is a random $\sqrt{N}\times\sqrt{N}$ matrix consisting of $M$ ones and $N-M$ zeros. The entries of $\Omega$ are not independent, but as $n$ increase, we can approximate $\Omega$ by using discrete independent identically distributed variables $\mathcal{P}\left(x=1\right)=M/N$ and $\mathcal{P}\left(x=0\right)=\left(N-M\right)/N$, which is a Bernoulli distribution with mean $q=M/N$ and variance $q\left(1-q\right)$. We denote this matrix $\Omega_{idd}$. We have
\begin{equation}
\rho_{\mathbb{A}_{N,M}} \sim \Omega_{idd}^{\dagger}\Omega_{idd}/M \, . 
\label{eq_rhoA_Hidd}
\end{equation}

\onecolumngrid
\begin{minipage}{0.98\textwidth}
    \begin{minipage}[b]{0.47\textwidth}
        \centering
        \begin{figure}[H]
            \includegraphics[scale=0.65]{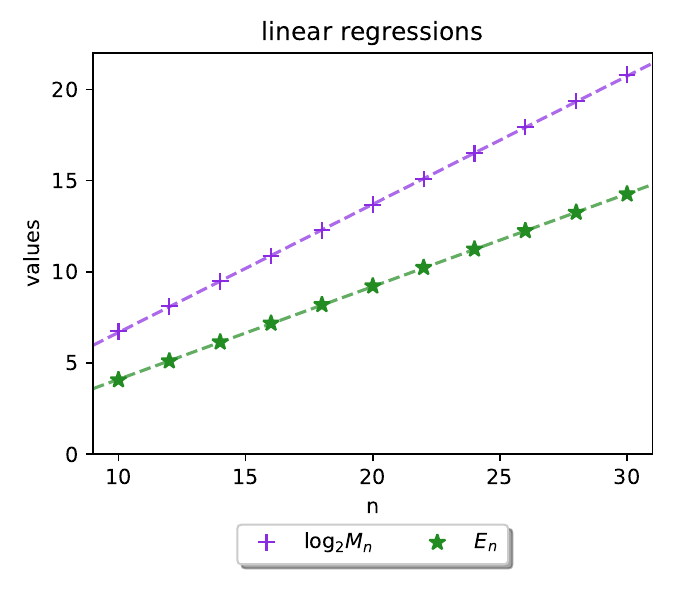}
            \caption{Linear regressions of $E_n$ and $\log_2{M_n}$. The dashed lines correspond to Eq.~\eqref{eq_linear}.}
            \label{fig_Linear}
        \end{figure}
    \end{minipage}
    \hfill
    \begin{minipage}[b]{0.47\textwidth}
        \centering
        \begin{figure}[H]
            \includegraphics[scale=0.65]{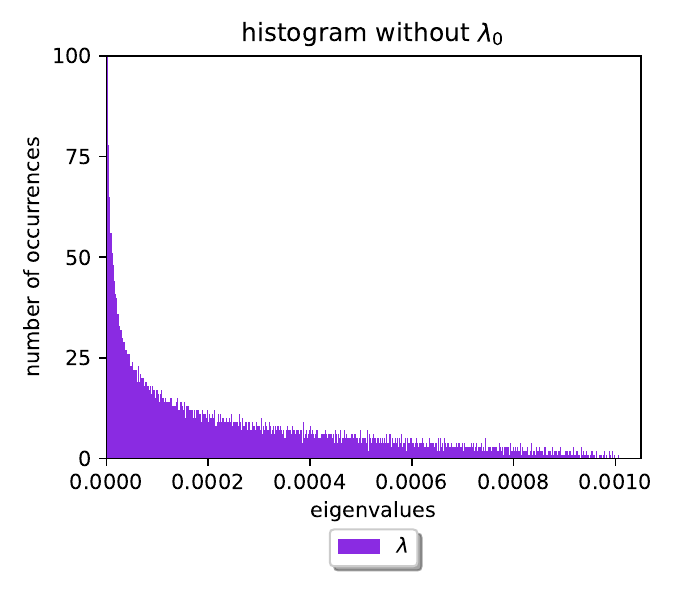}
            \caption{Histogram of the eigenvalues $\lambda$ (excluding $\lambda_0$) of the density matrix $\rho_{\mathbb{A}_{N, M}}$ for $n=24$ and $M=93018$.}
            \label{fig_Spectral}
        \end{figure}
    \end{minipage}
\end{minipage}
\vspace{0.1\columnsep}
\vspace{0.1cm}
\twocolumngrid

The Perron–Frobenius theorem \cite{meyer2023matrix} states that the largest eigenvalue for a non-negative real matrix is confined between the smallest sum of the row and the biggest sum of the row. From the relation in Eq. (\ref{eq_rhoA_Hidd}), we can deduce that the average entries of $\rho_{\mathbb{A}_{N,M}}$ are $M/N\sqrt{N}$. Therefore, the average sum of the row of matrix $\rho_{\mathbb{A}_{N, M}}$ is $M/N$. 

We have $\lambda_0=M/N$, using the following result from Ref. \cite{lin2024entanglement}, the average entropy (in natural logarithm) of a random reduced density matrix conditioned by the dominant eigenvalue $\lambda_0$ is given by
\begin{equation}
E_{\lambda_0}=\left(1-\lambda_0\right)\left(\ln{\alpha}-\ln\left(1-\lambda_0\right)-\frac{\alpha}{2\beta}\right)-\lambda_0\ln{\lambda_0} \, ,
\label{eq_E_alpha_beta_lambda}
\end{equation}
where $\alpha$ is the size of the subsystem and $\beta$ is the size of its complementary system. In this article $\alpha=\beta=\sqrt{N}$. We can have Eq. \eqref{eq_T_mn}, an analytical approximation of $E_{N, M}$. As shown in FIG. \ref{fig_Approximation}, $T_{N,M}$ reaches a maximum while $M=0$, the value is $n/2-c$, for more interpretation of this value, see Appendix \ref{app_partitions}. The leading term of $n/2$ in this expression aligns with Eq. \eqref{eq_linear}.
\begin{widetext}
\begin{equation}
T_{N,M}=\frac{\left(N-M\right)\left(\frac{3n}{2}-\log_2\left(N-M\right)-c\right)+M\left(n-\log_2 M\right)}{N} \, , \quad c = \frac{\log_2 e}{2} \simeq 0.721 \, .
\label{eq_T_mn}
\end{equation}
\end{widetext}

\section{Arithmetic sequences in the Momentum Space and $k$-almost primes}

Before diving into the result of this section, we first introduce some background of the example that will appear. In Ref. \cite{latorre2013quantum,latorre2014there,garcia2020prime}, the authors have studied the entanglement of the Prime state. The Prime state $\ket{\mathbb{P}_N}$ is defined as an equally weighted superposition of prime numbers in $n$ qubits computational basis where each prime number $p=p_0 2^0 + p_1 2^1 + ... + p_{n-1} 2^{n-1}$ is implemented as $\ket{p}=\ket{p_{n-1},...,p_{0}}$,

\begin{equation}
\ket{\mathbb{P}_N}=\frac{1}{\sqrt{\pi\left(N-1\right)}}\sum_{p \in\text{prime}}\ket{p} \, .
\end{equation}

The natural bi-partition entanglement entropy of the Prime state for even $n$ is investigated in Ref. \cite{latorre2014there}. In this example, we explore its other variant, referred to as the $k$-almost Prime state. $k$-almost Prime is a number that has $k$ prime factors, usually denoted as $p_k$. Here, we define their union as the equal superposition of $\bigcup_{i=1}^k p_k$
\begin{widetext}
\begin{equation}
\ket{\mathbb{U}_{N, 1}}=\ket{\mathbb{P}_N} \, , \quad \ket{\mathbb{U}_{N, k}}=\frac{1}{\sqrt{\pi_1\left(N-1\right)+...+\pi_k\left(N-1\right)}}\left(\sum_{p_1\in\text{$1$-almost prime}}\ket{p_1}+...+\sum_{p_k\in\text{$k$-almost prime}}\ket{p_k}\right) \, .
\end{equation}
\end{widetext}
where $\pi_k\left(N-1\right)$ is the amount of $k$-almost Prime numbers less than $N$. We note its entanglement
\begin{equation}
U_{N, K}=S_{\mathbb{U}_{N, k}} \, .
\end{equation}

There are two properties of $\ket{\mathbb{U}_{N, k}}$ that we should notice. First, $\mathbb{U}_{N, k}\subset\mathbb{U}_{N, k+1}$, the number of superpositions is increasing. When $k=1$, $M_n<\pi_1\left(N-1\right)$, so every $\ket{\mathbb{U}_{N, k}}$ located on the right side of the Maximally Entangled Sequences. When $k=n-1$, every computational basis except $\ket{0...0}$ are occupied. Secondly, unlike random sequences, $\mathbb{U}_{N, k}$ have an underlying structure inherited from prime numbers, which are highly unpredictable. Therefore, the entanglement of the Prime state is high, and as $k$ increases, the information about primes dissipates. These properties make $\ket{\mathbb{U}_{N, k}}$ a fair comparison to $\ket{\mathbb{A}_{N, M}}$.

In numerical simulation, we observe the preservation of entanglement of arithmetic sequences in the position and momentum space, which can be calculated by applying the Quantum Fourier Transform (QFT) to the state, as illustrated in FIG. \ref{fig_QFT}. This observation can not be extended to a theorem because we can easily find a counter-example, the entanglement before and after applying QFT to the Bell states (which also represent the sequence $\{0, 3\}$ for $n=2$) $\left(\ket{00}+\ket{11}\right)/\sqrt{2}$ are not identical. However, we can give an explanation to this non-trivial phenomenon, that QFT introduces small entanglement into the quantum state \cite{chen2023quantum}, which is also intuitive, the entropy in the position and momentum space should carry the same amount of information, which encodes the sequence itself.

Furthermore, arithmetic characteristics of number sequences can be visualized with the help of entanglement. In FIG. \ref{fig_QFT}, $U_{N, k}$ (slightly convex) decreases faster than $E_{N, M}$ (concave), verifying the second property of $\ket{\mathbb{U}_{N, k}}$, that the entanglement is lost, not only because the basis is populating; also the information about prime numbers is vanishing.

\onecolumngrid
\begin{minipage}{0.98\textwidth}
    \begin{minipage}[b]{0.47\textwidth}
        \centering
        \begin{figure}[H]
            \includegraphics[scale=0.65]{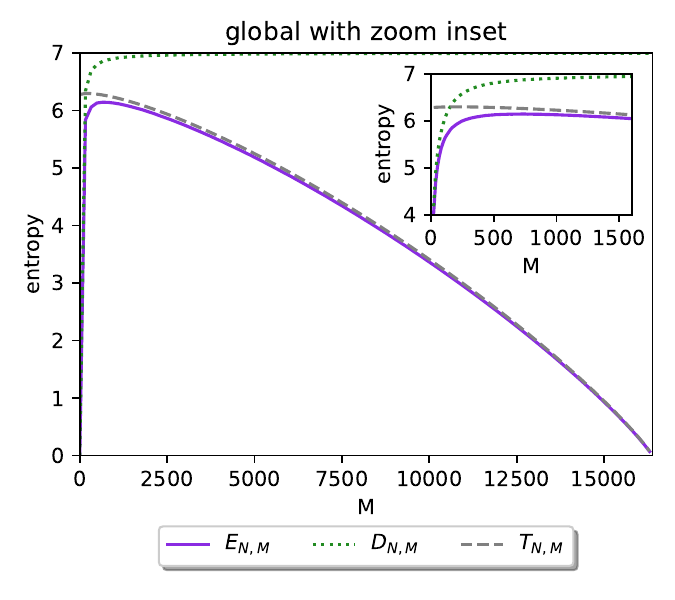}
            \caption{For $n=14$, comparison between the numerical $E_{N, M}$ and the analytical $T_{N, M}$. The approximation $D_{N, M}$ for $M \ll N$ is given in Appendix \ref{app_diagonal}.}
            \label{fig_Approximation}
        \end{figure}
    \end{minipage}
    \hfill
    \begin{minipage}[b]{0.47\textwidth}
        \centering
        \begin{figure}[H]
            \includegraphics[scale=0.65]{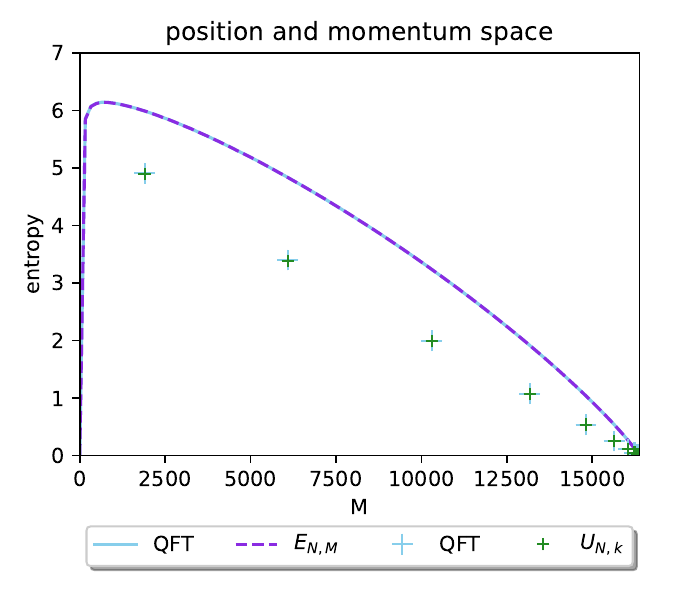}
            \caption{For $n=14$, the numerical results of $E_{N, M}$ and $U_{N,k}$. The entanglement before and after applying QFT is identical for arithmetic sequences.}
            \label{fig_QFT}
        \end{figure}
    \end{minipage}
\end{minipage}
\vspace{0.1\columnsep}
\vspace{0.2cm}
\twocolumngrid

\section{Conclusions}

In this article, we have proposed an encoding of any arithmetic sequence into a quantum state, allowing us to explore its entanglement properties. Given the highly random nature of most arithmetic sequences, their von Neumann entropy tends to concentrate along a specific trajectory, defined as the average entropy of the arithmetic sequence, for which we provide numerical and asymptotical analysis. 

Additionally, we have presented the example of unions of $k$-almost prime numbers to illustrate how this trajectory can serve as a benchmark for investigating arithmetic features of number series in a quantum way, both in position and momentum space.

Our work opens the door to further research on the entanglement properties of arithmetic sequences and their quantum analogs, potentially paving the way for new quantum algorithms and computational methods that leverage these unique characteristics. 

\begin{acknowledgements}

The authors extend their gratitude to S. Carrazza for valuable support in numerical simulation. 
GS acknowledges financial support through the Spanish MINECO grant PID2021-127726NB-I00, the CSIC Research Platform on Quantum Technologies PTI-001, and the QUANTUM ENIA project Quantum Spain through the RTRP-Next Generation within the framework of the Digital Spain 2026 Agenda.

\end{acknowledgements}

\FloatBarrier

\begin{appendices}

\section{R\'{e}nyi entropy}\label{app_renyi}

\begin{figure}
\centering
\includegraphics[scale=0.63]{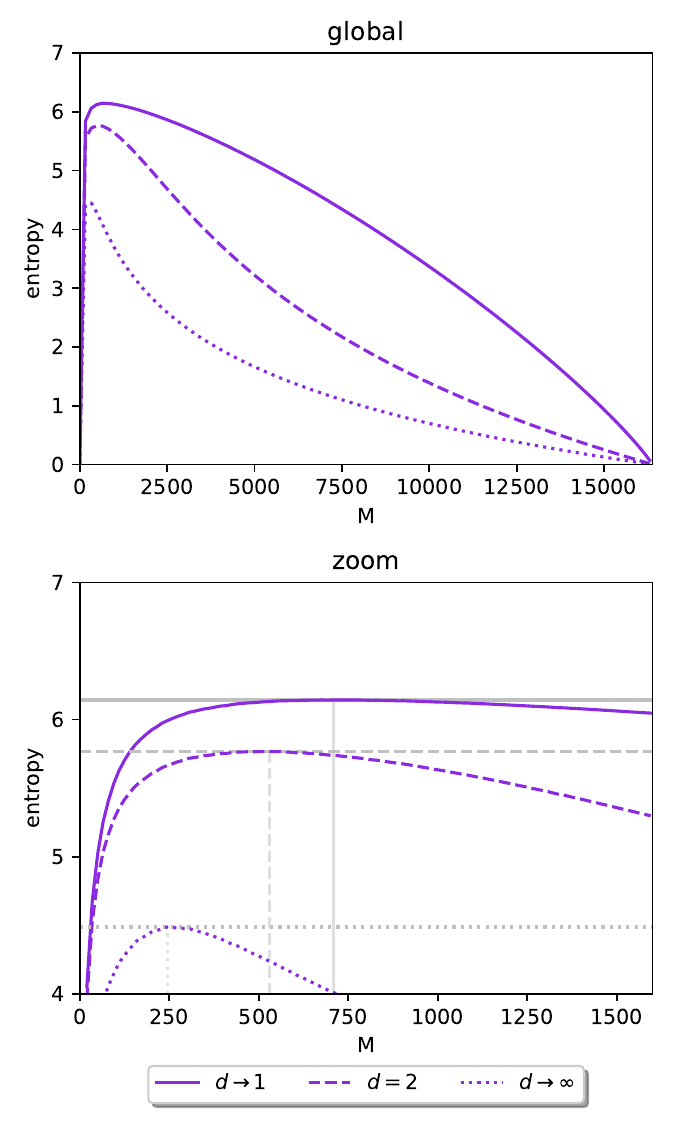}
\caption{For $n=14$, the entanglement of arithmetic sequences with varying orders $d$ of R\'{e}nyi entropy averaged over different states using the natural bi-partition. The plot includes their maximum values and the corresponding $M$.}
\label{fig_Renyi}
\end{figure}

The dependence of $E^{\left(d\right)}_{N, M}$ on $M$ for different orders $d$ is shown in FIG. \ref{fig_Renyi}, for R\'{e}nyi entropy defined as
\begin{equation}
E^{\left(d\right)}=\frac{1}{1-d}\log_2\left(\sum^{\sqrt{N}-1}_{j=0}\lambda_j^d\right) \, .
\end{equation}

\section{Partitions}\label{app_partitions}

The value $n/2-c$ is the average bi-partition entropy \cite{page1993average,bengtsson2017geometry,sen1996average} of a system of dimension $N$, which is also the entanglement of the Haar random states \cite{mele2024introduction}. One way to generate them is \cite{biswas2021inhibition}
\begin{equation}
\ket{\mathbb{H}_N}=\sum_{j=0}^{N-1} c_j \ket{j} \, ,
\label{eq_R_n}
\end{equation}
where $c_j$ are Gaussian random complex variables with mean $0$ and the standard deviation $1/\sqrt{N}$ for the state to be normalized. 

We take one sample of the Haar random state and one sample of $\ket{\mathbb{A}_{N, M_n}}$ and compare their entanglement over 1,000 bi-partitions. As shown in FIG. \ref{fig_Partition}, there is no overlap between these distributions, and $\ket{\mathbb{A}_{N, M_n}}$ exhibits more significant fluctuations across different partitions.

We employ the greedy algorithm to identify the best sample, defined as the quantum state with the highest entanglement across all partitions, using the same $M_n$. For 10 qubits, there are $\binom{10}{5}/2 = 126$ partitions, and for 14 qubits, there are $\binom{14}{7}/2 = 1,716$ partitions. The best results from 1,000,000 samples of $\ket{\mathbb{A}_{N, M_n}}$ with $n=10$, and 1,000 samples with $n=14$ are presented in FIG. \ref{fig_Greedy}.

\section{Analysis for $E_{N, M}$ as $M\ll N$}\label{app_diagonal}

When $M$ is small, the distribution of eigenvalues 
of the reduced density matrix associated with an arithmetic sequence is discrete, and $T_{N, M}$ does not provide a reliable approximation. Therefore, we need to explain the sparse-to-dense transition of $\rho_{\mathbb{A}_{N, M}}$. When it is sparse, the influence of the off-diagonal elements on the eigenvalues is negligible. The trace of the density matrix is $1$, and each element of the matrix is a multiplication of $1/M$.

We give an example to model the behavior of diagonal elements. Suppose we have $M$ little balls, each weight $1/M$, and $\sqrt{N}$ boxes. These balls are thrown randomly into boxes. We can use the mean of $D_{N, M}$ over many samples to simulate the entropy of a sparse $\rho_{\mathbb{A}_{N, M}}$
\begin{equation}
D_{N, M}=-\sqrt{N}\langle \frac{W}{M}\log_2\left(\frac{W}{M}\right)\rangle \, ,
\label{eq_G_mn}
\end{equation}
where $W$ is the number of balls in each box that follows the Bernoulli trial. For one particular box, the probability $q_W$ that it receives exactly $W$ balls is
\begin{equation}
\begin{aligned}
q_W
=\binom{M}{W}\left(\frac{1}{\sqrt{N}}\right)^W\left(\frac{\sqrt{N}-1}{\sqrt{N}}\right)^{M-W} \, .
\end{aligned}
\end{equation}

Finally, we have the expression
\begin{equation}
D_{N, M}=-\sqrt{N}\sum^{\sqrt{N}}_{W=1}q_W \frac{W}{M} \log_2\left(\frac{W}{M}\right) \, .
\end{equation}

As the number of balls $M$ increases, the weight inside each box will become evenly distributed, and the sum will converge towards $n/2$.

\onecolumngrid
\begin{minipage}{0.98\textwidth}
    \begin{minipage}[b]{0.47\textwidth}
        \centering
        \begin{figure}[H]
            \includegraphics[scale=0.61]{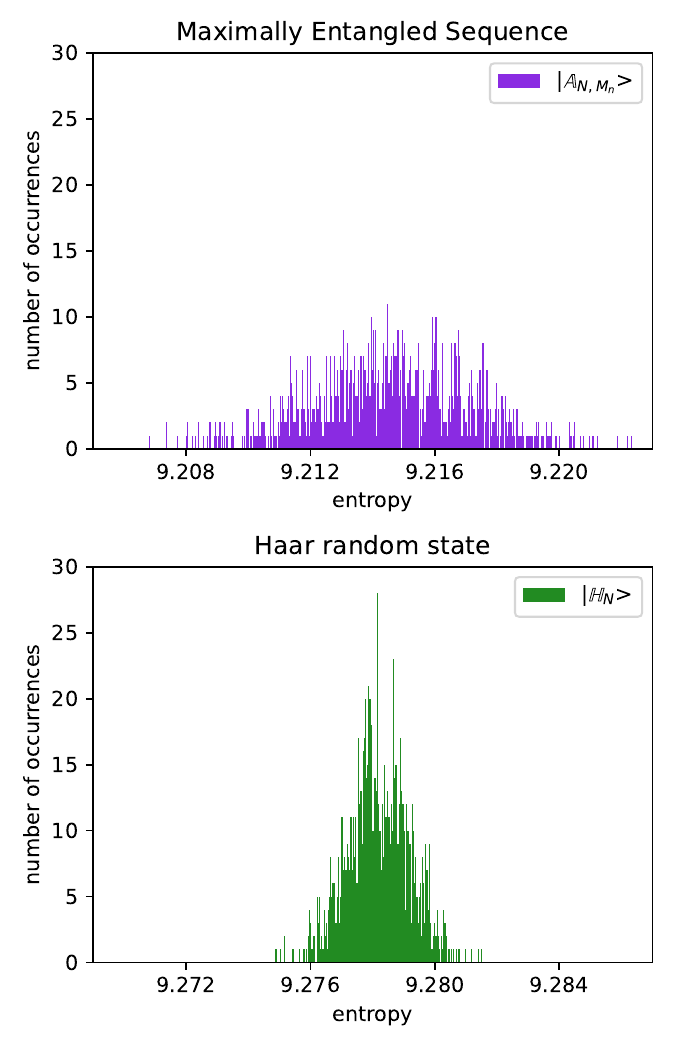}
            \caption{For $n=20$, histograms display the entanglement entropy of 1,000 partitions for one Maximally Entangled Sequence compared to one Haar random state.}
            \label{fig_Partition}
        \end{figure}
    \end{minipage}
    \hfill
    \begin{minipage}[b]{0.47\textwidth}
        \centering
        \begin{figure}[H]
            \includegraphics[scale=0.61]{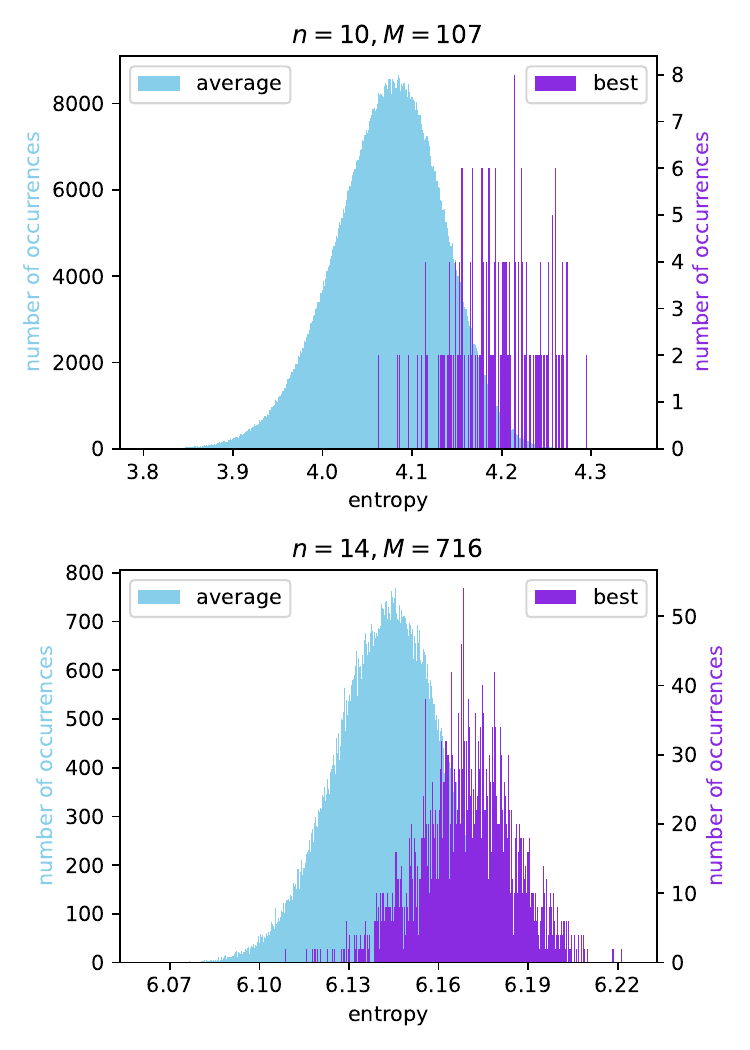}
            \caption{The comparison of entanglement of the best samples found in the greedy algorithm and the average $S_{\mathbb{A}_{N, M_n}}$ distribution for $n=10$ and $n=14$.}
            \label{fig_Greedy}
        \end{figure}
    \end{minipage}
\end{minipage}
\twocolumngrid

\end{appendices}

\end{document}